\documentclass[12pt]{iopart} 
\usepackage{graphicx}
\usepackage{amssymb}
\begin{document} 
\title[Transfer matrix solution of the WSME model augmented by
short range interactions]
{Transfer matrix solution of the
Wako-Sait{\^o}-Mu\~noz-Eaton model augmented by arbitrary
short range interactions}
\author{V I Tokar$^{1,2}$ and H Dreyss\'e$^1$}
\address{$^1$IPCMS, UdS--CNRS, UMR 7504, 23 rue du Loess,
F-67034 Strasbourg, France}
\address{$^2$Institute of Magnetism, National Academy of Sciences,
36-b Vernadsky Boulevard, 03142 Kiev-142, Ukraine}
\ead{tokar@ipcms.u-strasbg.fr}
\begin{abstract}
The Wako-Sait{\^o}-Mu\~noz-Eaton (WSME) model, initially introduced in
the theory of protein folding, has also been used in modeling the RNA
folding and some epitaxial phenomena.  The advantage of this model is
that it admits exact solution in the general inhomogeneous case
(Bruscolini and Pelizzola, 2002) which facilitates the study of
realistic systems.  However, a shortcoming of the model is that it
accounts only for interactions within continuous stretches of native
bonds or atomic chains while neglecting interstretch (interchain)
interactions.  But due to the biopolymer (atomic chain) flexibility,
the monomers (atoms) separated by several non-native bonds along the
sequence can become closely spaced. This produces their strong
interaction.  The inclusion of non-WSME interactions into the model
makes the model more realistic and improves its performance.  In this
study we add arbitrary interactions of finite range and solve the new
model by means of the transfer matrix technique.  We can therefore
exactly account for the interactions which in proteomics are
classified as medium- and moderately long-range ones.
\end{abstract}
\maketitle
\section{Introduction} 
The WSME model is a generalization of the one dimensional (1D) lattice
gas model with nearest neighbor (NN) interatomic pair interactions.  In
addition to the NN interactions, cluster interactions are present
inside continuous chains of adjasent atoms.  The model was initially
introduced by Wako and Sait{\^o} \cite{wako_saito1,wako_saito2} and by
Mu\~noz and Eaton \cite{ME,ME_ea98} to understand protein folding.  The
role of atoms played the peptide bonds.  Recently this model was used
to describe RNA folding \cite{RNA_ME}.  Furthermore, a similar model
was derived in a theory of strained epitaxy \cite{PRB,j_phys}.

The physical meaning of the cluster interactions is easily
understandable in the case of coherent strained epitaxy.  Let us assume
that besides the attractive NN interaction $v_1<0$ the interatomic
potential has a rigid core which does not let the atoms approach each
other closer than the core diameter $d$ \cite{philMag}.  So if the
diameter is larger than the substrate lattice spacing $a$, the adatoms
within an atomic chain will be displaced from the centers of the
deposition sites by $u_j\propto f$.  The requirement of
coherence means that the displacements should be small in order for the
displaced atoms remained within the same lattice cell.  In general this
condition will be violated for sufficiently long chains but in the
present study we consider only finite systems and assume that the
misfit is sufficiently small for the condition of coherence to be
satisfied. In this case the misfit energy of atom $j$ in the harmonic
approximation can be estimated as $ku_j^2/2$, where $k$ is the
curvature of the substrate potential near its minimum.  The 
atomic displacements within a chain of length $l$ can be found from 
symmetry considerations as
\begin{equation} 
\label{ eq0}
u_{\pm j}=\pm\left\{\begin{array}{llr}
f(j+1/2) & j=0,1,\dots,l/2-1& l\mbox{ even}\\
fj&j=0,1,\dots,(l-1)/2&l\mbox{ odd}
\end{array}\right..
\end{equation} 
With the use of identities
$$\sum_{j=1}^mj=m(m+1)/2$$
and
$$\sum_{j=1}^mj^2=m(m+1)(2m+1)/6$$
the total energy of the chain of length $l$ 
after some algebra can be calculated as
\begin{equation} 
\label{ El0}
E^{(l)}=V^{(1)}l+v_1(l-1)+(kf^2/24)l(l^2-1),
\end{equation} 
where $V^{(1)}$ is the adsorption energy per atom.  

Let us assume that $N_a<N$ adatoms are gathered onto $N_a/l$ equal
chains of length $l$.  The total energy of the system in this case will
be equal to $(N_a/l)E^{(l)}$.  Thus, the energy minimum at fixed $N_a$
will coincide with the minimum of $E^{(l)}/l$.  From \eref{ El0} it is
easy to see that such a minimum always exists provided $f\neq0$.  This
produces a simple model of self-assembly of size calibrated coherent
nanostructures similar to quantum dots.

Formally the hamiltonian (or, more precisely, the configuration
dependent free energy) of the WSME model is 
\cite{wako_saito1,wako_saito2,ME,ME_ea98,PRB,j_phys,pelizzola}
\begin{equation} 
\label{ H0}
H^{(N)}_{\rm WSME}=\sum_{l=1}^N\sum_{i=l}^N V_i^{(l)} \prod_{k=i-l+1}^i n_k,
\end{equation} 
where in the case of epitaxy $N$ is the total number of deposition sites, 
$n_j=0,1$ describes the occupation of site $j$ by the gas atom,
$V_i^{(l)}$ are inhomogeneous (i.\ e., site-dependent) interactions
within the continuous atomic chains of length $l$ ending at site $i$,
as can be seen from the product in \eref{ H0}.  As was shown in
Equation (5) of \cite{PRE}, if chain energies $E^{(l)}$ for all $l$ are
known, the values of $V^{(l)}$ in \eref{ H0} can be found as the
discrete second derivative of $E^{(l)}$ with respect to $l$.
Furthermore, to simplify notation we assume the chemical potential
$\mu$ to be included (with the minus sign) into the parameters 
$V_i^{(1)}$.

In the case of biopolymers the interpretation of the interactions in
the model is different. Firstly, the ``atoms'' in this case are either
the amino-acid residues \cite{wako_saito1,wako_saito2} or the covalent
bonds \cite{ME,ME_ea98,pelizzola} which are assumed to be present in
two states:  the native and the non-native one corresponding to the
values 1 and 0 of a binary variable, respectively
\cite{wako_saito1,wako_saito2,ME,ME_ea98,RNA_ME}.  Thus, $N$ can be
either the number of peptide bonds connecting $N+1$ amino-acid residues
or the number of the residues themselves.  In the more developed bond
model $V_i^{(1)}$ is the loss of conformation entropy by the bond in
the process of formation of the native state
\cite{ME,ME_ea98,pelizzola}.  Cluster interactions $V^{(l)}$ in \eref{
H0} can be presented in the form \cite{pelizzola}
\begin{equation}
V_i^{(l)}=\varepsilon_{ji}\Delta_{ji},
\end{equation}•
where $i$ and $j=i-l+1$ are the 1D coordinates of two peptide bonds;
$\Delta_{ji}=1$ if the bonds are in contact with each other and is
equal to zero otherwise; $\varepsilon_{ji}$ is the inter-residue
interaction energy between residues $i$ and $j+1$. Farther details 
are given in \cite{pelizzola}.  

The binary matrix $\Delta_{ji}$ defines the contact map of a protein in
its native state.  It depends on the definition of the residue
contact.  The major parameter here is the cutoff distance between atoms
which separates the atoms considered to be in contact from remote
atoms.  For example, if the distance is chosen to be 8 \AA\ then each
residue  on average contacts with approximately ten other residues (see
Table 1 in \cite{interact_review}).  For smaller values of the cutoff
chosen in \cite{wako_saito2} the average number of contacts per residue
is $\lesssim3$ (see Table I in \cite{wako_saito2}).

The qualitative similarity between the epitaxial and the folding models
can be seen on the lattice protein folding model considered in
\cite{3Dlattice}.  According to the model, the folding starts at random
places in the process of nucleation of a local native structure.  The
binary bond variables inside the regions are all equal to unity while
in other regions the variables are zero.  Statistically such behavior
can be described by the WSME model.

Despite being 1D, the WSME model has two peculiarities hampering its
exact solution.  The first is the absence of the translational
invariance which makes inapplicable the efficient techniques of the
homogeneous case \cite{wako_saito1,wako_saito2,PRB,PRE}.  The other
peculiarity is that the hamiltonian \eref{ H0} contains long-range
interactions so the conventional transfer matrix (TM) method cannot be
used.  Because of these peculiarities, the exact solution for the WSME
model at equilibrium in the inhomogeneous case was found only recently
\cite{pelizzola}.  This solution, on the one hand, greatly facilitates
the study of the kinetics of folding \cite{zamparo_pelizzola}, on the
other hand, it allows for the modeling of the strained epitaxy in
inhomogeneous environments, such as alloyed substrates \cite{j_phys}.
This latter case is of interest in connection with engineering
applications where the 1D nanostructures (such as nanowires,
nanomagnets, nanotubes, etc.) may have important applications
\cite{nature_review,nanowires0,nanowires,zigzag_wire,magneticreview,%
exp,ONsteps08,Co/Au2,scienceCNT1,scienceCNT2,Kcnt60,cnt_ground_state07}.
Because in the device environment the wire (for example) may traverse
different chemical surroundings, make turns, experience disordered
substrate potential due to doping, etc., the interaction parameters
describing the model should in general be position-dependent.

In the epitaxial systems, however, the inhomogeneous WSME model
describes only 1D chains of atoms or molecules.  But in practical
applications more than monatomic structures can be needed in order,
e.\ g., to enhance the conductivity of a nanowire or to increase the
magnetic moment of a nanomagnet.  These structures may consist of
several adjacent atomic rows on a terrace of a vicinal surface or on
the surface of a nanotube.  Such quasi-2D structures can be described
with the use of the WSME model only if at least further neighbor pair
interactions are added to the hamiltonian \eref{ H0}.

\begin{figure} 
\includegraphics[viewport = -20 480 300 650]{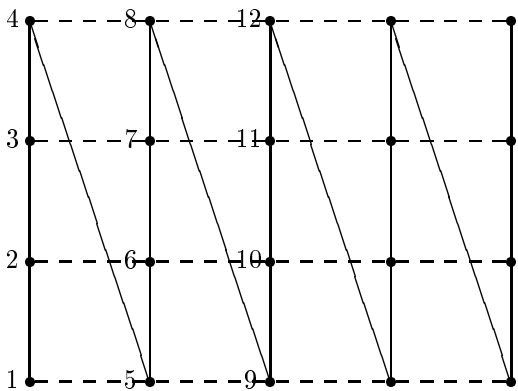}
\caption{\label{Fig1}Black points represent the deposition sites on the
substrate; solid line with numbers shows a possible mapping of the
sites on a 1D lattice with integer coordinates. Dashed lines connect
nearest neighbor sites on the substrate lattice with rectangular
geometry.  These sites are the nearest neighbors on the substrate but
the fourth neighbors in the 1D lattice coordinates.  For farther information
see the text.} 
\end{figure}

Indeed, let us consider the topology of the deposition sites shown on
\fref{Fig1}. This topology may correspond to the deposition sites on
the terrace of a vicinal surface with a rectangular geometry. In this
case atoms 2 and 6 or 7 and 11 will be nearest neighbors on the
substrate lattice but not along the 1D lattice.  But if the substrate has the
geometry of triangular lattice with the angle 2--1--5 being equal to
$120^{\circ}$, the atoms 1--6 and 7--12 (for example) will constitute
additional nearest neighbor pairs.  If, farther, the sites in
\fref{Fig1} are rolled into a cylinder with the chiral vector (4,0),
the pairs of sites of type 1 and 4 or 9 and 12 become nearest neighbors
too.  Other tube topologies will bring together other atoms.  An
example of the nanotube with chirality (4,1) will be given in section
\ref{example}. Obviously in all these cases the neglect of the
interactions between the atoms on the nearest neighbor sites of the
substrate lattice will qualitatively change the physics of the system
under consideration.  But such interactions in the 1D lattice
coordinates will be further neighbor interactions (4-th neighbors in
the above case of the tube of rectangular geometry) which do not enter
into the WSME hamiltonian \eref{ H0}.

Similar arguments can be applied also to 2- and 3D lattice models of
proteins \cite{nnn+3B,3Dlattice}.  Mu\~noz and Eaton noted in this
connection \cite{ME} that the inclusion of interactions between the
bonds belonging to different native stretches into the WSME model not
only will improve quantitative description of the kinetics but also
``would add considerable flexibility to possible structural mechanisms
by producing additional routes between the denatured and native
states''.

The non-WSME interactions may be of qualitative importance in
differentiating between the proteins and the RNA folding.  While in
some respects being similar, the folding of the two polymer types
differ \cite{RNA_proteins}.  In particular, in contrast to proteins,
the RNA native state is hierarchical in that the secondary structure is
energetically well separated from the tertiary one and can be
considered as a collection of base pairings
\cite{RNA2nd,nucl_phys,RNA_TM}.  Because of this, in the standard model
of the RNA the energetics is governed by the medium and long range pair
interactions \cite{nucl_phys,bund_hwa}.

Last but not least, even when the larger distance along the 1D lattice
corresponds to larger separation in real space so that further neighbor
interactions are small, in some cases they also may cause qualitative
phenomena.  For example, if the model considered at the beginning of
this section is augmented by the substrate mediated repulsive dipole
pair interactions \cite{lau_kohn,marchenko_parshin2} or, alternatively,
by attractive interaction of some other origin \cite{PRE} the model, in
addition to the self-assembly and size calibration, will simulate the
important phenomenon of self-organization of quantum dots into periodic
arrays.  Thus, there exist many situations when farther neighbor
interactions are among the most important ones in the system and cannot
be neglected without qualitatively changing the system properties.

But besides the pair interactions, sufficiently strong cluster
interactions of non-WSME kind are also usually present in epitaxial
systems (see the next section).  In this study we present therefore a
TM solution of the WSME model augmented by arbitrary short range
interactions.  Because of the restrictions imposed by the TM technique,
in practice the method will be restricted to the interactions of
relatively short range.  Nonetheless, fairly large radii up to those
which in proteomics are classified as the medium- and moderately
long-range interactions are feasible to exact treatment within our
approach.

In the next section we introduce our extension of the WSME model, in
section \ref{solution} explain its formal solution and in section
\ref{example} present a simple illustrative calculation.  In the
concluding section \ref{discussion} we discuss our results.
\section{Extended WSME model} 
The pair interactions, such as the Coulomb or the van der Waals are the
most ubiquitous in nature and their account should be the primary goal
in extending the WSME model.  But more complex cluster interactions
(CIs) can also be important, especially in metallic systems where the
pair approximation holds only approximately. For example, {\em ab
initio} calculations show that interactions within atomic trios
deposited at the surface may have the same magnitude as the nearest
neighbor pair interactions, i.\ e., to be among the strongest in the
system \cite{trio_interactions2,arXiv09}.  To account for all possible
CIs, in the {\em ab initio} theories of alloys and epitaxial systems
the method of interatomic interaction expansion over a complete set of
CIs has been developed \cite{ECI,ducastelle,trio_interactions2}.  It is
pertinent to note that binary alloy is formally equivalent to the
lattice gas model.  In order for our approach to be compatible with
this powerful technique, we developed it for an arbitrary set of CIs
restricted only by the maximum values of their interaction radii.  This
is necessary for the computational tractability of the TM equations.

Let us first consider the most general hamiltonian which includes
all possible CIs in the system of size $N$
\begin{equation} 
\label{ HG}
H^{(N)}_{N-1}=\sum_{C=1}^{2^N-1} W_C n_{N}^{c_{N-1}}n_{N-1}^{c_{N-2}}
\dots n_2^{c_1}n_1^{c_0},
\end{equation} 
where the subscript $N-1$ denotes the maximum interaction radius.  We
define the radius of a CI as $r=i_{max}-i_{min}$, where $i_{max(min)}$
is the largest (smallest) index of $n_i$ in the cluster, $c_i=0,1$
and, by definition, $n_i^0\equiv1$.  The CIs in \eref{ HG} are
characterized by sequences of binary digits which can be gathered into
the number
\begin{equation} 
\label{ barC}
\bar{C}=(c_{N-1}\dots c_1c_0)_B, 
\end{equation} 
where the bar over a number denotes that its binary representation is
meant; the subscript $B$ denotes that the term within parentheses is
the binary representation, not the product and $W_C$ is the strength of
the corresponding CI.  In \eref{ HG}, \eref{ H0} and below we list the
terms in the products in reverse order because in our TM approach it is
convenient to number the sites from right to left.

The total number of CIs in hamiltonian \eref{ HG} is of $\Or(2^N)$
which is a huge number for even modest systems of sizes $N\approx50$
characteristic for the smallest proteins.  Only a small part of
$\Or(N^2)$ of the CIs from \eref{ HG} enters into the hamiltonian
\eref{ H0}.  Obviously, in the general inhomogeneous case it would be
impossible to take into account all interactions for a system of
practical interest.  To make the problem manageable, we restrict the
extent of the interactions by some maximum radius $R$.

The extended WSME model we will solve in the next section has the
hamiltonian
\begin{equation} 
\label{ H1}
H^{(N)}=H^{(N)}_{\rm WSME}+H^{(N)}_{R},
\end{equation} 
where the second term on the right hand side is defined as the sum of
all such terms in \eref{ HG} that do not contain interactions of radii
exceeding $R$ and, besides, in order to avoid double counting these
terms should not enter into \eref{ H0}.
\section{\label{solution}Recursive transfer matrix solution}
In physical terms the main difference between the models represented by
hamiltonians \eref{ H0} and \eref{ H1} is as follows.  In  \eref{ H0}
due to the specific form of interactions the energy of any
configuration is the sum of energies of the continuous atomic chains
(or the stretches of the peptide bonds) it contains.  The interactions
in \eref{ H0} become zero as long as atomic clusters are separated by a
single empty site.  Thus, in the homogeneous case the system can be
considered as a mixture of non-interacting molecules of $N$ kinds
(different sizes) \cite{ PRE}.  Presumably because of this simplicity
the homogeneous case was solved much earlier than the general case
\cite{wako_saito1,wako_saito2}.

The additional term in \eref{ H1} changes drastically the situation
even in the homogeneous case because now not only different chains
interact but their interactions are quite nontrivial.  For example, in
the case of $R=14$ considered in \cite{arXiv09}, up to eight islands
may be interacting via appropriate CIs.  Because of this, there seems
to be no way of accounting for all possible situations except through
their direct enumeration. In the case of finite range interactions and
in 1D this can be done recursively by adding sites to the system one by
one.

So let us for the time being neglect in \eref{ H0} all interactions
whose radii exceed $R$.  Because of the finite interaction range, when
adding a site to the system consisting of $K\geq R$ sites only the
interactions with the last $R$ sites need be taken into account.  The
accounting can be done with the use of the vector partition function
$\vec{Z}^{(K)}$ whose components are the partial traces over all except
the last $R$ sites
\begin{equation} 
\label{ Z_partial}
Z^{(K)}_{n_K,n_{K-1},\dots,n_{K-R+1}}
=\Tr_{n_1,n_2,\dots,n_{K-R}} \exp(-H^{(K)}).   
\end{equation} 
Here $H^{(K)}$ is a hamiltonian \eref{ H1} for a $K$-site system which
contains only interactions within the range not exceeding $R$. The
total partition function is found from \eref{ Z_partial} as
\begin{equation} 
\label{ Z_K}
Z^{(K)}=\sum_{\bar{\alpha}={\bar{0}}}^{\overline{2^R-1}}
Z^{(K)}_{\bar{\alpha}},
\end{equation}
where the bar over the number has the same meaning as in \eref{ barC}.

As was shown in Appendix of \cite{arXiv09} and will be explained in
more detail below, a recurrence relation for $\vec{Z}^{(N)}$ in the
number of sites in the system $N$ can be established.  This technique
is an extension of the methods developed in connection with the 2D
Ising model in \cite{screwed_cilinder,screwed_cilinder2}.  Its
advantage is that it deals with sparse TMs which provide considerable
gain in computational effort in the case of large $R$.

If we assume that the vector partition function for the system of size
$N-1$ is known then the partition function for the size $N$ can be
calculated recursively with the use of the sparse TM as
\begin{equation} 
\fl
\label{ RR}
\left(\!\!\!
\begin{array}{c}
\circ\circ\dots\circ\circ\\
\circ\circ\dots\circ\bullet\\
\vdots\\
\circ\bullet\dots\bullet\bullet\\
\bullet\circ\dots\circ\circ\\
\bullet\circ\dots\circ\bullet\\
\vdots\\
\bullet\bullet\dots\bullet\bullet
\end{array}
\!\!\right)^{(N)}\!\!\!=
\left(
\begin{array}{cccccccc}
1&1&0&0&\dots&0&0&0\\
0&0&1&1&\dots&0&0&0\\
\vdots&\vdots&\vdots&\vdots&\ddots&\vdots&\vdots&\vdots\\ 
0&0&0&0&\dots&0&1&1\\
b_{\bar{0}}&b_{\bar{1}}&0&0&\dots&0&0&0\\
0&0&b_{\bar{2}}&b_{\bar{3}}&\dots&0&0&0\\
\vdots&\vdots&\vdots&\vdots&\ddots&\vdots&\vdots&\vdots\\ 
0&0&0&0&\dots&0&b_{\overline{2^R-2}}&b_{\overline{2^R-1}}
\end{array}
\right)_{N}
\!\left(\!\!\!
\begin{array}{c}
\circ\circ\dots\circ\circ\\
\circ\circ\dots\circ\bullet\\
\vdots\\
\circ\bullet\dots\bullet\bullet\\
\bullet\circ\dots\circ\circ\\
\bullet\circ\dots\circ\bullet\\
\vdots\\
\bullet\bullet\dots\bullet\bullet
\end{array}
\!\!\right)^{(N-1)},
\end{equation} 
where the column vectors correspond to $\vec{Z}^{(N)[(N-1)]}$, the empty
and filled circles describe the empty ($n_i=0$) or filled ($n_i=1$)
sites in the subscripts of the partial partition functions in \eref{
Z_partial} and the subscript $N$ of the TM is the site index for all
$b_{\bar{\alpha}}$ entering the matrix.  We note that we use the same
symbol $N$ for the system size and for the recurrent relation to
stress that at every iteration we obtain the (vector) partition
function of a system corresponding to some size $N$.

In the case of finite-range interactions the structure of TM in \eref{
RR} is easily understood.  Having added site $N$ to the system
consisting of $N-1$ sites we first have to account for the interaction
of this site with the rest of the system and then take the trace over
the ($N-R$)-th site because with the radius of interactions being $R$
all interactions of this site with the rest of the system have already
been taken into account.  Taking the trace amounts to adding with
appropriate weights two $Z^{(N-1)}$ differing by the filling of site
$N-R$.  In the case of the empty site $N$ the weights are equal to
unity because the empty site does not interact with anything.  These
terms occupy the upper half of the TM \eref{ RR}.  The lower half of
the matrix contains the terms corresponding to the interaction of the
{\em occupied} site $N$ with the rest of the system.  The term
\begin{equation} 
\label{ b_i}
b_{\bar{\alpha}N}=\exp(-\Delta E_{\bar{\alpha}N}/k_BT) 
\end{equation} 
is the Boltzmann weight corresponding to the interaction of the atom at
site $N$  with the configuration of atoms corresponding to
$Z^{(N-1)}_{\bar{\alpha}}$; $\Delta E_{\bar{\alpha}N}$ in \eref{ b_i}
is the energy of interaction of the atom with configuration
$\bar{\alpha}$.

Now, what have to be changed in order to include the arbitrary range
interactions of the WSME type into the recursion scheme \eref{ RR}? It
turns out that only the last equation need be modified.  This is
because the hamiltonian \eref{ H0} contains only the interactions
inside continuous  chains. But all components of the state vector
$\vec{Z}^{(N-1)}$ except the last one contain at least one empty site
among the last $R$ sites.  Therefore, the extent of the chain
interactions is restricted by the distance to the nearest empty site
and thus is smaller than $R$.

In the last component, however, all sites are filled. So when adding an
additional $N$-th site filled with an atom we do not know which
interactions of the WSME type should be taken into account as the last
$R$ atom may belong to a chain of any length---from $R$ to $N-1$.  We
overcome this difficulty in a straightforward manner by simply taking
into account all the possibilities. Namely, we replace  the last
two-term equation in the set \eref{ RR} with the sum over all
configurations where the last $R$ sites belong to a chain of length
greater or equal to $R$. This can be achieved with the use of the
component ${Z}^{(M)}_{\overline{2^R-2}}\equiv
\bullet\bullet\dots\bullet\bullet\circ$ with all sites except the first
one being filled. Note that the positions are counted from right to
left.  When adding chains of different lengths to this component we can
control the total chain length and thus know which interactions from
\eref{ H0} should be taken into account.

Formally this is done as follows.  In the course of the recursive
solution we keep the array of components ${Z}^{(M)}_{\overline{2^R-2}}$
for $M=R-1,\dots,N-1$ (the explanation of the term $R-1$ is given
below);  in another array we gather the chain energies $E_N^{(l)}$.
These account for all interatomic interactions entering \eref{ H1}
{\em inside} the chains of length $l$ ending at site $N$.  The chains
are assumed to be isolated so no interchain interactions enter 
$E_N^{(l)}$.  By attaching a chain of length $N-M+R-1$ to the configuration
corresponding to ${Z}^{(M)}_{\overline{2^R-2}}$ which amounts to
multiplying the latter by the corresponding Boltzmann factor, we obtain a
configuration with a continuous chain of atoms starting on site $M-R+1$
and ending on site $N$.  As is seen, the $(R-1)$-atom chain in
${Z}^{(M)}_{\overline{2^R-2}}$ ending at site $M$ and the chain ending
at $N$ overlap at sites inside the (sub)chain of length $R-1$.  In the
equation below this double counting is taken care of by the division by
the necessary Boltzmann factor corresponding to the chain of length
$(R-1)$:
\begin{equation} 
\label{ Z15}
Z^{(N)}_{\overline{2^R-1}}=\sum_{M=R-1}^{N-1}
\exp[-E_N^{(N-M+R-1)}/k_BT] \tilde{Z}^{(M)}_{\overline{2^R-2}},
\end{equation}
where
\begin{equation} 
\label{ Z14a}
\tilde{Z}^{(M)}_{\overline{2^R-2}}={Z}^{(M)}_{\overline{2^R-2}}
/\exp[-E_M^{(R-1)}/k_BT].
\end{equation}  
The meaning of \eref{ Z15} is simple: the component of $\vec{Z}^{(N)}$
with the last $R$ sites being filled is obtained as the sum of all
possible configurations having the chains of length $R\le l\le N$ as
their end sites.  As is easy to see, the factor
$\tilde{Z}^{(M)}_{\overline{2^R-2}}$ is sufficient for accounting for
all short-range interactions of the chain of length $N-M+R-1$ with the
rest of the system because the atoms at sites $M+1$ and larger cannot
reach the atoms beyond $M-R$ due to the finite interaction range.  The
only remaining problem is connected with the longest chain of length
$N$ which should comprise the whole system because
${Z}^{(R)}_{\overline{2^R-2}}$ starts with an empty site.  This
difficulty is overcome by initializing the recurrence \eref{ RR} with
$\vec{Z}^{(R-1)}$, i.\ e., with the system containing only $R-1$ sites
instead of $R$.  The fillings of these sites correspond to the last
$R-1$ sites in the vectors in \eref{ RR}, i.\ e., the rightmost column
in these vectors should be crossed out so the component
${Z}^{(R-1)}_{\overline{2^R-2}}$ has all its sites filled. The
components corresponding to the empty crossed out sites are calculated
as the conventional Boltzmann factors while in the cases when the
omitted site was filled they are all set to zero. The validity of this
initialization of the recurrence \eref{ RR} can be proven either by a
straightforward calculation of $\vec{Z}^{(R)}$ via one iteration step
and comparing it with $\vec{Z}^{(R)}$ calculated straightforwardly, or
by associating with the crossed out column a fictitious 0-th site which
has an infinite on-site energy. Thus, on the one hand, the initial
$\vec{Z}^{(R-1)}$ corresponds to the system of size $R$
($0,\dots,R-1$); on the other hand, the components corresponding to
this site being filled are all zero, as suggested above.
\section{\label{example}Illustrative calculation} 
As can be seen from \eref{ Z15}, formally our algorithm is quadratic in
the system size $N$.  This means that for sufficiently large systems
the calculations may become prohibitively difficult to perform.  The
sizes of biopolymers met in nature, however, are restricted
\cite{pdb}.  Because from practical point of view of major interest are
natural biological molecules, we will restrict our discussion to this
case.  A typical protein consists from about 500 amino acid residues
\cite{bund_hwa}.  So in order to assess numerical performance of our
algorithm we consider for simplicity an epitaxial model of this size.
The epitaxial systems of similar sizes are of interest also for the
nanoengineering.  Because the devices of sizes in tens of nanometers
(hundreds of atoms) are efficiently modeled in the framework of
continuum approximations \cite{cont_approx,cont_approxB,cont_approxC},
our approach may be useful in studies of smaller few-nanometer
structures \cite{nature_review}.  Thus, the length in 500 atomic
diameters (about 100 nm) is, presumably, an upper limit of interest for
the atomic simulations of epitaxial systems (the issue of their width
will be discussed below).

We consider coherent strained epitaxy on the surface of a finite size
screw (4,1) nanotube with rectangular substrate lattice geometry (see
\fref{Fig1}) with homogeneous interactions and consisting of 500
deposition sites.  This geometry was chosen because the diameter four
would correspond, {\em inter alia}, to a model of the $\alpha$-helix
similar to that considered in \cite{epl} but with additional pair
interactions.  This may be used to model the helices with different
interbond interactions to better describe their properties.  The (4,1)
topology means that sites 2 and 6 or 7 and 11 in \fref{Fig1} are
nearest neighbors along the direction parallel to the tube axis while
the sites along the solid line are all equivalent.  For example, the
interaction between atoms at sites 2 and 3 is the same as between those
at sites 8 and 9 because all points along the line belong to a helix.
The potential of the substrate (the tube surface) is periodically
corrugated along the helix and will be treated in the harmonic
approximation \eref{ El0}, as discussed in the Introduction.

Besides the positive misfit energy, the atoms in our model experience,
apart from the NN interaction $v_1$, small attraction between the first
and the second neighbors along the helix.  Because of the homogeneity
of the model, the nomenclature of \eref{ HG} is superfluous, so below
we denote this interaction as $v_2$. Besides, $v_4$ will designate the
repulsion between the atoms which are the fourth neighbors along the
helix but are NN on the substrate surface (see \fref{Fig1}).  This
model qualitatively describes the large misfit systems studied in
\cite{cnt_ground_state07} and \cite{Kcnt60}.  In these papers it was
found that while on the tubes of large diameters the interaction
between the nearest neighbor adatoms is repulsive along both
directions, on those of small diameters the interaction along the high
curvature direction became attractive due to the increased interatomic
distance.  But the interaction in the direction of small curvature
along the tube axis remains repulsive even for the small-diameter
tubes.

In the explicit calculations below we used the following values in units
of the NN attractive interaction $v_1<0$:
\begin{equation} 
\label{ v_i}
v_2=0.3v_1,\quad v_4=0.2|v_1|\quad \mbox{and}\quad 
kf^2\approx6.8\cdot10^{-3}|v_1|.
\end{equation} 
The energy of an isolated chain of length $l$ can be obtained from
\eref{ El0} by adding to it the terms due to $v_2$ and $v_4$
\begin{equation} 
\label{ E_ch}
E^{(l)}=\sum_{i=1,2,4}(l-i)_{>0}v_i+(kf^2/24)l(l^2-1)-\mu l,
\end{equation} 
where the subscript ${>0}$ means that only positive values of $(l-i)$
contribute and $V^{(1)}$ in \eref{ El0} was set equal to $-\mu$.

\begin{figure} 
\includegraphics[viewport = -50 210 200 420, scale = 0.9]{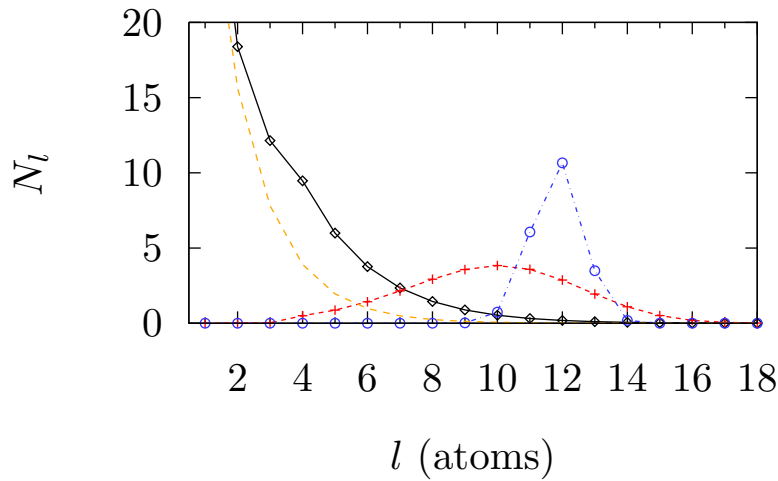}
\caption{\label{Fig2}Low temperature size calibration of self-assembled
atomic clusters in the model described in the text.  The number of
clusters of different sizes at different temperatures:
\opendiamond---$T=|v_1|$, +---$T=0.1|v_1|$ and
\opencircle---$T=0.01|v_1|$. The curves are guides to the eye except
the monotonous dashed curve which describes random coverage.}
\end{figure}

Numerical values of the pair interactions \eref{ v_i} were chosen in
such a way that in the absence of misfit ($f=0$) the reduced energy
$E^{(l)}/l$ did not have a global minimum at finite value of $l$ so
that the system were of phase separation type.  This means that the
atoms at low temperature tend to gather into one cluster.  In the
presence of the misfit, however, $E^{(l)}/l$ has a local minimum at
$l=12$.  This choice means that the chain makes three turns around the
tube which approximately corresponds to the structure of the typical
$\alpha$-helix.  This model was solved with the recursive technique of the
previous section at three different temperatures for the system
consisting of 500 sites at half coverage (250 atoms).  In \fref{Fig2}
are shown the size distributions of chains of different lengths on the
surface of a cylinder under consideration.  As is seen, at the highest
temperature the size distribution is similar to the random distribution
of atoms while at the lowest temperature it exhibits very good size
calibration with $\gtrsim96\%$ of atoms belonging to chains of lengths
11, 12 or 13.  Thus, in the presence of the misfit the atoms gather
into chains of about 12 atoms each.  This result is in accord with the
theory \cite{lannoo} but it was not obvious from the start because
the interisland interactions are known to shift the calibrated size
from its noninteracting value at the minimum of $E^{(l)}/l$.

All calculations were performed on a modern Intel\textregistered\
processor with the use of Python scripts.  This choice was motivated by
the problem of numerical underflow or overflow which appeared in the
calculations.  The problem is rooted in the exponential scaling of the
partition function with the system size:  ${Z}^{(N)}\sim\exp(N\phi)$,
where $\phi$ is the reduced (per site) grand potential.  Because of
this, at sufficiently large $N$ ${Z}^{(N)}$ may acquire arbitrarily
large or small values which exhaust any fixed numerical range.  In
Python libraries, however, there exists the module {\tt decimal} which
allows for calculations with extremely small and/or extremely large
numbers.

When present, this problem severely hampers the calculations.  For
example, in the above model with $N=500$ and the parameters shown in
\fref{Fig2} calculation of ${Z}^{(500)}$ required only a fraction of a
second because the conventional double precision arithmetic was
sufficient:  The maximum values of ${Z}^{(500)}$ were of
$\Or(10^{101}$).  But this means that the system with, e.\ g., 5000
sites will have $Z^{(5000)}\sim\Or(1010)$ which in our approach would
require the use of the module {\tt decimal}.  Indeed, explicit
calculation gave $Z^{(5000)}\approx1.5\cdot10^{1011}$.  This
calculation took almost 35 minutes which is about 150 times longer than
the calculation at this size without the module.  Repeated with the
parameters where the double precision was sufficient it took only about
14 seconds.  Thus, the software realization of high-precision
arithmetics costs more than two orders of magnitude in performance.  If
this is characteristic for all such software, much better choice is to
use the quadruple precision realized in some C/C++ and Fortran
compilers.  In addition to much smaller overhead due to higher
precision, the compiled languages offer additional speed up in about
two orders of magnitude in comparison with the interpreted languages
such as Python.  In this way the calculations with large biopolymers
should be very fast.  For example, the calculation of $Z^{(35000)}$
with the Python script in the case when the double precision was
sufficient took less than 12 minutes which with the compiled language
should take only a few seconds.

It should be reminded that all calculations were performed for $R=4$.
As can be seen from \eref{ RR} and \eref{ Z15}, the equations have the
form of scalar products of vectors of sizes $2^R$ and $N$,
respectively.  This means that the calculations are trivially
parallelizable, so the execution speed at large $R$ will depend on the
number of processors available for the calculation and on the
performance of one processor.  The latter can be assessed from the
model calculation of $Z^{(500)}$ with $R=20$ or $\Or(10^6)$ equations
in the set \eref{ RR} with the use of a Python script which lasted
about 15 minutes.  This means that with a compiled language the time of
the calculation will measure in seconds.  We estimate that the
calculations with $R$ in the range $\lesssim40$ should be feasible on
modern supercomputers. 
\section{\label{discussion}Discussion}
In this paper we presented a transfer matrix solution of the WSME model
extended to account for arbitrary short range interactions.  The
transfer matrix approach is, in principle, a universal technique
capable of solving any lattice problem with short range interactions.
In practice, however, it is restricted to relatively small interaction
radii due to the exponential growth of the computational effort with
$R$. This restriction does not allow the method to be considered as a
universal tool for obtaining exact solutions in dimensions D$>$1
because in statistical mechanics one is usually interested in the
thermodynamic limit which corresponds to $R\to\infty$ and thus is
unaccessible to the TM technique in truly 2- or 3D systems
\cite{screwed_cilinder,screwed_cilinder2,arXiv09}.

The natural biopolymers, however, though sometimes very large, are
restricted in their maximum size, so their characteristic dimensions
are also finite.  According to current nomenclature the radii of short
and medium range interactions in proteins do not exceed 20 residues
\cite{interact_review}.  As we saw, this case causes no difficulty even
for a single processor computer.  On a supercomputer with tens to
hundreds parallel processors even the moderately long-range
interactions of the extent $R\lesssim40$ studied in \cite{3Dlattice}
should cause no problems.  Thus, in the case of proteins our approach
potentially allows for the {\em exact} solution of the extended WSME
model with arbitrary medium- and moderately long-range interactions.
According to \cite{interact_review} (see their fig. 4), the
interactions with ranges exceeding 40 constitute only about 5\% of all
interactions.  Thus, the technique developed in the present paper
allows to improve up to 95\% of all interactions.

The RNA molecules are less amenable to the study within our TM
technique because the double-stranded nature
\cite{RNA_proteins,bund_hwa} of the polymer makes the pair interactions
very long-ranged, up to the total molecule length when the first and
the $N$-th nucleotides pair.  Therefore, the pair interactions in the
ranges extending not farther than about 20 pairs along the stem away
from the hairpin loops can be treated within our approach.  The pseudo
knots formed by nearby hairpin heads are other potential candidates for
the description with non-WSME interactions \cite{nucl_phys}.  An
alternative way of describing the nucleotide interactions is to include
them into the WSME part \cite{RNA_ME}.

In the case of epitaxial systems the maximum interaction radius
$R\lesssim40$ lattice units restricts the application of the method to
the nanotubes of similar and even lesser circumference \cite{arXiv09}.
In this connection it is pertinent to note that $R\approx20$
approximately corresponds to the upper limit of the tube size when the
high curvature of small diameter nanotubes can qualitatively change the
ordering of adsorbates consisting from large atoms
\cite{cnt_ground_state07}. In the case of deposition on the terraces
the upper limit of width ($\lesssim40$ atomic rows) even exceeds the
width of the terraces (16 rows) used in \cite{Co/Au2} for epitaxial
growth of magnetic nanostructures.  Such a restriction of the
accessible widths is not very serious from the practical point of
view.  In the case of wide nanowires in tens of atoms the atomic
resolution is not very important because an error in a few atoms can be
neglected in most cases.  Nanostructures of such sizes can be
efficiently simulated within continuum approximations
\cite{cont_approx,cont_approxB,cont_approxC}.

Furthermore, the short range interactions can be used to describe the
substrate propagated elastic dipole-dipole interaction in 1D model of
strained epitaxy proposed in \cite{PRB}.  The dipole-dipole interaction
behaves as the inverse cube of the distance
\cite{lau_kohn,marchenko_parshin2} and so at $R\sim10-20$ in 1D systems
can be neglected in most cases.

It should be mentioned that the direct push interaction \cite{philMag}
leading to the interactions of the WSME type in 1D or on the screw
tubes is not operative in wires on the terraces of width greater than
two (on the non-rectangular substrate the wires consisting of two rows
may still contain such a contribution).  In this case the origin of
some of the WSME type interactions can be different. For example, the
interaction corresponding to the largest cluster containing all atoms
differentiates two cases: the fully filled terrace and the terrace with
one vacancy.  Such an interaction may account for the volume
contribution to the vacancy formation enthalpy.  Thus, there is enough
interesting epitaxial systems (and we mentioned only a few of them)
which can be simulated with the exact transfer matrix technique
developed in the present paper.  
\ack
The authors acknowledge CNRS for support of their collaboration.  One
of the authors (V.I.T.) expresses his gratitude to Universit{\'e}
de Strasbourg and IPCMS for their hospitality.  We thank M. Alouani
for a critical reading of the manuscript.


\section*{References}
\begin{thebibliography}{10}
\providecommand{\url}[1]{\texttt{#1}}
\providecommand{\urlprefix}{URL }
\providecommand{\eprint}[2][]{\url{#2}}

\bibitem{wako_saito1}
Wako H and Sait{\^o} N 1978 \emph{J. Phys. Soc. Jpn.} \textbf{44} 1931

\bibitem{wako_saito2}
Wako H and Sait{\^o} N 1978 \emph{J. Phys. Soc. Jpn.} \textbf{44} 1939

\bibitem{ME}
Mu{\~n}oz V and Eaton W~A 1999 \emph{Proc. Natl. Acad. Sci.} \textbf{96} 11311

\bibitem{ME_ea98}
Mu{\~n}oz V, Henry E~R, Hofrichter J and Eaton W~A 1998 \emph{Proc. Natl. Acad.
  Sci.} \textbf{95} 5872

\bibitem{RNA_ME}
Imparato A, Pelizzola A and Zamparo M 2009 \emph{Phys. Rev. Lett.} \textbf{103}
  188102

\bibitem{PRB}
Tokar V~I and Dreyss{\'e} H 2003 \emph{Phys. Rev. B} \textbf{68} 195419

\bibitem{j_phys}
Tokar V~I and Dreyss\'e H 2004 \emph{J. Phys.: Condens. Matter} \textbf{16}
  S2203

\bibitem{philMag}
Tokar V~I and Dreyss\'e H 2008 \emph{Phil. Mag.} \textbf{88} 2747

\bibitem{pelizzola}
Bruscolini P and Pelizzola A 2002 \emph{Phys. Rev. Lett.} \textbf{88} 258101

\bibitem{PRE}
Tokar V~I and Dreyss{\'e} H 2003 \emph{Phys. Rev. E} \textbf{68} 011601

\bibitem{interact_review}
Gromiha M~M and Selvaraj S 2004 \emph{Prog. Biophys. Mol. Biol.} \textbf{86}
  235

\bibitem{3Dlattice}
Abe H and Wako H 2009 \emph{Physica A} \textbf{388} 3442

\bibitem{zamparo_pelizzola}
Zamparo M and Pelizzola A 2006 \emph{Phys. Rev. Lett.} \textbf{97} 068106

\bibitem{nature_review}
Barth J~V, Costantini G and Kern K 2005 \emph{Nature} \textbf{437} 671

\bibitem{nanowires0}
Bowler D~R 2004 \emph{J. Phys.: Condenc. Matter.} \textbf{16} R721

\bibitem{nanowires}
Owen J~H~G, Miki K and Bowler D~R 2006 \emph{J. Mater. Sci.} \textbf{41} 4568

\bibitem{zigzag_wire}
Gonz{\`a}lez C, Snijders P~C, Ortega J, P{\'e}rez R, Flores F, Rogge S and
  Weitering H~H 2004 \emph{Phys. Rev. Lett.} \textbf{93} 126106

\bibitem{magneticreview}
Himpsel F~J, Ortega J~E, Mankey G~J and Willis R~F 1998 \emph{Adv. Phys}
  \textbf{47} 511

\bibitem{exp}
Gambardella P, Brune H, Kern K and Marchenko V~I 2006 \emph{Phys. Rev. B}
  \textbf{73} 245425

\bibitem{ONsteps08}
Negulyaev N~N, Stepanyuk V~S, Hergert W, Bruno P and Kirschner J 2008
  \emph{Phys. Rev. B} \textbf{77} 085430

\bibitem{Co/Au2}
Repain V, Baudot G, Ellmer H and Rousset S 2002 \emph{Europhys. Lett.}
  \textbf{58} 730

\bibitem{scienceCNT1}
Zhou C, Kong J, Yenilmez E and Dai H 2000 \emph{Science} \textbf{290} 1552

\bibitem{scienceCNT2}
Wang Z, Wei J, Morse P, Dash J~G, Vilches O~E and Cobden D~H 2010
  \emph{Science} \textbf{327} 552

\bibitem{Kcnt60}
Yang X and Ni J 2004 \emph{Phys. Rev. B} \textbf{69} 125419

\bibitem{cnt_ground_state07}
Lueking A~D and Cole M~W 2007 \emph{Phys. Rev. B} \textbf{75} 095425

\bibitem{nnn+3B}
Salvi G and {P De Los Rios} 2003 \emph{Phys. Rev. Lett.} \textbf{91} 258102

\bibitem{RNA_proteins}
Thirumalai D and Hyeon C 2005 \emph{Biochemistry} \textbf{44} 4957

\bibitem{RNA2nd}
Schuster P, Fontana W, Stadler P~F and Hofacker I~L 1994 \emph{Proc. R. Soc.
  Lond. B} \textbf{255} 279

\bibitem{nucl_phys}
Orland H and Zee A 2002 \emph{Nucl. Phys. B} \textbf{620[FS]} 456

\bibitem{RNA_TM}
Wolfsheimer S, Burghardt B, Mann A and Hartmann A~K 2008 \emph{J. Stat. Mech.}
  P03005

\bibitem{bund_hwa}
Bundschuh R and Hwa T 1999 \emph{Phys. Rev. Lett.} \textbf{83} 1479

\bibitem{lau_kohn}
Lau K~H and Kohn W 1977 \emph{Surf. Sci.} \textbf{65} 607

\bibitem{marchenko_parshin2}
Marchenko V~I and Parshin A~Y 1980 \emph{Sov. Phys. JETP} \textbf{52} 129

\bibitem{trio_interactions2}
Luo W and Fichthorn K~A 2005 \emph{Phys. Rev. B} \textbf{72} 115433

\bibitem{arXiv09}
Tokar V~I and Dreyss\'e H 2009 \emph{Preprint}  arXiv:0912.2680v2

\bibitem{ECI}
Sanchez J~M, Ducastelle F and Gratias D 1984 \emph{Physica} \textbf{128A} 334

\bibitem{ducastelle}
Ducastelle F 1991 \emph{Order and Phase Stability in Alloys}.
\newblock North-Holland, Amsterdam

\bibitem{screwed_cilinder}
Kramers H~A and Wannier G~H 1941 \emph{Phys. Rev.} \textbf{60} 252

\bibitem{screwed_cilinder2}
Domb C 1949 \emph{Proc. Royal Soc. Series A} \textbf{196} 36

\bibitem{pdb}
  \urlprefix\url{http://www.pdb.org}

\bibitem{cont_approx}
Tu Y and Tersoff J 2007 \emph{Phys. Rev. Lett.} \textbf{98} 096103

\bibitem{cont_approxB}
Grima R, DeGraffenreid J and Venables J~A 2007 \emph{Phys. Rev. B} \textbf{76}
  233405

\bibitem{cont_approxC}
Eggleston J~J and Voorhees P~W 2002 \emph{Appl. Phys. Lett.} \textbf{80} 306

\bibitem{epl}
Flammini A, Banavar J~R and Maritan A 2002 \emph{Europhys. Lett.} \textbf{58}
  623

\bibitem{lannoo}
Priester C and Lannoo M 1995 \emph{Phys. Rev. Lett.} \textbf{75} 93

\end{thebibliography}
\end{document}